\documentstyle[emulateapj,psfig]{article}
\def\gs{\mathrel{\raise0.35ex\hbox{$\scriptstyle >$}\kern-0.6em 
\lower0.40ex\hbox{{$\scriptstyle \sim$}}}}
\def\ls{\mathrel{\raise0.35ex\hbox{$\scriptstyle <$}\kern-0.6em 
\lower0.40ex\hbox{{$\scriptstyle \sim$}}}}
\newcommand{\msun}{\odot}

\begin{document}

\title{Evidence for tidal stripping of dark matter halos in massive 
cluster-lenses}

\author{Priyamvada Natarajan$^{1}$, Jean-Paul Kneib$^{2}$ \& Ian Smail$^{3}$}
\affil{1 Department of Astronomy, Yale University, New Haven, CT, USA}
\affil{2 Observatoire Midi-Pyrenees, 14 Av. E.Belin, 31400 Toulouse,  
France}
\affil{3 Department of Physics, University of Durham, South Road, 
Durham DH1 3LE, UK}

\begin{abstract}
In this letter, we present the results of our study of galaxy-galaxy
lensing in massive cluster-lenses spanning $z = 0.17$ to $0.58$,
utilizing high-quality archival {\it Hubble Space Telescope} ({\it
HST}\,) data. Local anisotropies in the shear maps are assumed to
arise from dark matter substructure within these clusters. Associating
the substructure with bright early-type cluster galaxies, we quantify
the properties of typical $L^*$ cluster members in a statistical
fashion.  The fraction of total mass associated with individual
galaxies within the inner regions of these clusters ranges from
10--20\% implying that the bulk of the dark matter in massive lensing
clusters is smoothly distributed.  Looking at the properties of the
cluster galaxies, we find strong evidence ($>3$-$\sigma$ significance)
that a fiducial early-type $L^\ast$ galaxy in these clusters has a
mass distribution that is tidally truncated compared to equivalent
luminosity galaxies in the field. In fact, we exclude field galaxy
scale dark halos for these cluster early-types at $>10$-$\sigma$
significance.  We compare the tidal radii obtained from this lensing
analysis with the central density of the cluster potentials and find a
correlation which is in excellent agreement with theoretical
expectations of tidal truncation: $\log [r_t*] \propto (-0.6\pm 0.2)
\log [\rho_0]$.
\end{abstract}


\section{Introduction}

Galaxy-galaxy lensing provides a powerful tool to statistically
measure the mass and the details of the mass distribution for field
galaxies (Tyson et al.\ 1984; Brainerd et al.\ 1996). These studies
confirm the existence of massive dark matter halos around typical
field galaxies, extending to beyond 100\,kpc\footnote{We adopt
h=H$_o$/100km\,s$^{-1}$\,Mpc$^{-1}$=0.5 and $q_o=0.5$, $\Omega_o =
1$. Our results however, are not sensitive to values of the cosmological
parameters.}(Brainerd et al.\ 1996; Fischer et al.\ 2000; Smith et al.\ 2001b;
McKay et al.\ 2001).  The same technique can be modified and
implemented within clusters to constrain the masses of cluster
galaxies (Natarajan \& Kneib 1997, NK97; Geiger \& Schneider
1998). Successful application of the same to the rich, lensing cluster
AC\,114 at $z = 0.31$, suggests that the average $M/L$ ratio and
spatial extents of the dark matter halos associated with early-type
galaxies in such dense environments may differ from those of
comparable luminosity field galaxies (Natarajan et al.\ 1998, NKSE98).

The technique applied by NKSE98 quantifies the local weak distortions
in the observed shear field of massive cluster-lenses, as
perturbations arising from the massive halos of cluster galaxies (for
details see NK97).  By associating these perturbations with bright
early-type cluster members, the relative mass fraction in their halos
is constrained using a combined $\chi^2$-maximum likelihood
method. The strength of this approach is the simultaneous use of
constraints from the observed strong and weak lensing features.

The fractional mass in clusters associated with individual galaxy
halos has important consequences for the frequency and nature of
interactions (Moore et al.\ 1996; Ghigna et al.\ 1998; Okamato \& Habe
1999; Merritt 1983). The theoretical expectation is that the global
tidal field of a massive, dense cluster potential well should be
strong enough to truncate the dark matter halos of galaxies that
traverse the cluster core. In this letter, we test this expectation
using well calibrated mass models for rich clusters at $z\sim
0.17$--0.58 that utilize the observed strong lensing features --
positions, magnitudes, geometry of multiple images and measured
spectroscopic redshifts as well as the shear field.

\section{Galaxy-galaxy lensing in clusters}

The clusters in our study are modeled as a super-position of a smooth large-scale
potential and smaller scale potentials that are associated with bright
early-type cluster members: $\phi_{\rm tot} = \phi_{\rm clus} +
\Sigma_i \,\phi_{\rm p_i},$ where $\phi_{\rm clus}$ is the potential
of the smooth component and $\phi_{\rm p_i}$ are the potentials of the
perturbers (galaxy halos).  The resultant reduced shear $g$, \footnote{The
reduced shear $g$, a complex number, is defined in terms of the
convergence $\kappa$ and the shear $\gamma$, as $g =
{{\gamma}/{(1-\kappa)}}$, can be directly related to the measured
shape parameter $\tau = 2g/(1-g^{*}g)$}, $g$, and can in turn be
decomposed into contributions from the smooth clump and perturbers.

To quantify the lensing distortion induced by $\phi_{\rm tot}$,
the smooth cluster piece and individual galaxy-scale halos are
modeled self-similarly using the pseudo-isothermal elliptical 
(PIEMD) surface density profile, $\Sigma(R)$ derived by 
Kassiola \& Kovner (1993),
with a core-radius $r_0$, a truncation radius $r_t\,\gg\, r_0$ 
and an ellipticity
$\epsilon$. These parameters are tuned for both the smooth component
and the perturbers to obtain mass distributions on the relevant scales
(for details see \S2.2 of NK97). Note that $r_t$ characterizes the 
scale over which the local potential dominates. 
Additionally, assuming light traces
mass, the light distribution of the early-type cluster members is
related to the mass model via a set of the usual adopted scaling 
laws motivated by observations (see Brainerd et al. 1996):
\begin{eqnarray}
{\sigma_0}\,=\,{\sigma_{*}}({L \over L_*})^{1 \over 4};\,\,
{r_0}\,=\,{r_{0*}}{({L \over L_*}) ^{1 \over 2}};\,\,
{r_t}\,=\,{r_{t*}}{({L \over L_*})^{1 \over 2}}.
\end{eqnarray}
The effect of these assumed scaling laws on the maximum-likelihood
analysis has been explored in detail using simulations in NK97.
The total mass $M$ and the total mass-to-light ratio $M/L$ of cluster
galaxies then scales as follows:
\begin{eqnarray}
M\,\propto\,{\sigma_{*}^2}{r_{t*}}\,({L \over L_*})\,;\,\,{\frac{M}{L}}\,\propto\,
{\sigma_{*}^2}\,{r_{t*}}.
\end{eqnarray}
The ellipticity of the mass clumps is set to the
observed ellipticity of each cluster galaxy. Since, we are sensitive
to the induced local shear arising due to the integrated mass within
$r_t$ for the clumps, we are not sensitive to details of the 
assumed radial fall-off and the assumed inner slope of the profile
in the galaxy model. 

The measured local shear signal in the cluster at the
location of the $i$-th perturber is boosted in value due to two
effects (i) the contribution from the smooth cluster component --
$\gamma_c$ in the numerator, and (ii) the ($\kappa_c+\kappa_{p_i}$)
term in the denominator, which is positive and non-negligible in the
central cluster region. By taking into account a smooth
cluster model it is possible to reduce the error in the measured shear
from the cluster members:
\begin{eqnarray}
\nonumber
\left<{g_I}-{g_c}\right> = 
\left<{\gamma_{p_i}\over 1-\kappa_c -\kappa_{p_i}}\right>,\,\,
\left(\sigma^2_{g_I- g_c}\right) = {\sigma^2_{g_S} \over 2}\,\approx\, 
{\sigma^2_{p(\tau_S)}\over N_{bg}},
\end{eqnarray}
where $g_I$,$g_S$ and $g_c$ are the shapes of the lensed image (I), 
unlensed source (S) and the shear of the smooth cluster piece; 
${\sigma^2_{p(\tau_S)}} \sim 0.3$ is the width of the intrinsic 
ellipticity distribution of the sources, and $N_{bg}$ the number of 
background galaxies averaged over. Therefore, knowledge of a well 
calibrated strong lensing model makes the study of galaxy-galaxy 
lensing in clusters very viable.

A maximum-likelihood method is used to obtain significance bounds on
fiducial parameters that characterize a typical $L^\ast$ halo in the
cluster. We have extended the formalism developed in NK97 to include
the strong lensing data for the inner regions of the clusters, these
are used to obtain the best model $\chi^2$ fit followed by a
likelihood method that incorporates the constraints from the smoothed
shear field.

The likelihood function of the estimated probability distribution of
the source ellipticities is maximized for a set of model parameters,
given a functional form of the intrinsic ellipticity distribution
measured for faint galaxies. The entire inversion procedure to solve
the lensing equation is performed for each cluster within the {\sc
lenstool} utilities\footnote{A fully analytical ray-tracing routine
developed by Kneib (1993)}. Using a well-determined `strong lensing'
model for the inner-regions of the clusters along with the averaged
shear field and assuming a known functional form for $p(\tau_{S})$
from the field, the likelihood for a guessed model is computed using:
$\tau_{S_j} \,=\,\tau_{\rm obs_j}\,-{\Sigma_i^{N_c}}\,{\gamma_{p_i}}\,-\,
 \gamma_{c}$ for each faint galaxy $j$ where $\Sigma_{i}^{N_{c}}\,
{\gamma_{p_i}}$ is the sum of the shear contribution from $N_{c}$ 
perturbers, and $\gamma_{c}$ is the shear induced by the smooth component.
\begin{eqnarray}
 {\cal L}({{\sigma_{*}}},{r_{t*}}) = \Pi_j^{N_{gal}} p(\tau_{S_j}).
\end{eqnarray}
We compute ${\cal L}$ assigning the median redshift corresponding to
the observed source magnitude for each arclet (details of this
procedure can be found in NK97). The best fitting model parameters are
then obtained by maximizing the log-likelihood function $l$ with
respect to the chosen set of model parameters (${\sigma_{*}},
{r_{t*}}$) and the cluster parameters are simultaneously matched in an
iterative way.

\section{The HST cluster-lenses}  

For our analysis we select clusters at $z>0.1$ for which deep,
high-quality {\it HST} imaging is available and which contain
spectroscopically-confirmed multiply-imaged high redshift galaxies.
These lensed features are essential to construct a detailed mass
distribution for the cluster cores (e.g.\ Kneib et al.\ 1996; Smith et al.\
2001a), while the existence of spectroscopic redshifts allows us to
calibrate these mass distributions onto an absolute scale.  This
selection yields five clusters with redshifts
spanning $z=0.17$--0.58 for our analysis: A\,2218, 
A\,2390,  Cl\,2244$-$02,  Cl\,0024+16, 
and Cl\,0054$-$27.  In addition to these five clusters, we
also include our previous analysis of the $z=0.31$ cluster AC\,114
(NKSE98).  Details of the cluster properties are given
in Table~1 -- clearly these clusters do not
constitute a well-defined sample, for example their X-ray 
luminosities span an order of magnitude and their central mass
densities show a similarly large dispersion.  It is this latter property
which is of most interest for our analysis -- and the large range
spanned by the sample therefore provides a powerful test of the
variation in characteristics of galaxy halos with local environment.

The five new clusters were observed with {\it WFPC2}, four through 
the F814W ($I$) passband, with A\,2218 imaged in F702W
($R$), the total exposure times are listed in Table~1.
These typically comprise stacks of multiple
single-orbit (2.1--2.7\,ks) exposures, each of which is spatially
offset by an integer number of {\it WFC} pixels to allow the removal
of cosmic-rays and other artifacts.  The data were pipeline processed
by STScI and combined using standard {\sc stsdas} and {\sc iraf}
routines. For more details of the reduction see:
A\,2218,  Kneib et al.\ (1996); A\,2390, Pell\'o et al.\ (1999);
Cl\,0024+16 and Cl\,0054$-$27, Smail et al.\ (1997).
These fields reach typical 5-$\sigma$ detection limits for point
sources of $I\sim 26$ or $R\sim 26.5$.

Object catalogs and attendent shape information for faint galaxies are
obtained using the {\sc SExtractor} package (Bertin \& Arnouts 1996)
and detection criteria of 12 {\it WFC} pixels above the $\mu_R$ or
$\mu_I=25.0$\,mag\,arcsec$^{-2}$ isophote after convolution with a
0.3$''$ diameter top-hat filter.  Selection of the background galaxies
in these images employs a simple magnitude cut as previously
used by NKSE98, either $R=23$--26 or
$I=22.5$--25.5 -- this provides catalogs of typically $\sim 300$--400
faint galaxies in each field.  The magnitude limits employed should
select field galaxies at median redshifts of $z\sim 1$, well beyond
the clusters.  Visual morphological classifications of the brighter
galaxies in these fields, used to select the early-type cluster
galaxies, comes from Smail et al.\ (1997, 2001).

In addition to the strong lens model of AC\,114 employed by NKSE98,
lens models for three of the other clusters have been published
previously: A\,2218, Kneib et al.\ (1996); A\,2390, Pell\'o et al.\
(1999); Cl\,0024+16, Smail et al.\ (1996).  For the remaining two
lenses Cl\,2244$-$02 and Cl\,0054$-$27, we construct well constrained
mass models. The mass distribution in Cl\,2244$-$02 has significant
substructure, with the main clump centered around the brightest
cluster galaxy with a velocity dispersion of $600 \pm 80$ kms$^{-1}$, an
ellipticity of $0.17 \pm 0.1$ and the secondary clump offset about 
20 arcseconds away with a velocity dispersion of $300 \pm 40$
kms$^{-1}$ and an ellipticity of $0.1 \pm 0.05$. The cluster
Cl\,0054$-$27 by contrast has a smooth, but extended
mass distribution (with a cut-off radius of about $900 \pm 60$ kpc) with 
a central velocity dispersion of ${1100 \pm 100}$ kms$^{-1}$ and an 
ellipticity of ${0.1 \pm 0.05}$.

\section{RESULTS}

Before discussing the results, we describe the various tests we have
employed to confirm their reliability.
In order to check the authenticity of the signal the following null
tests were performed in the likelihood analysis for every cluster: (i)
randomizing the positions of the background galaxies, (ii) randomizing
the orientations of the background galaxies, and (iii) scrambling the
positions of cluster galaxies. No significant maxima were obtained in
the likelihood prescription for any of these tests in any of the
clusters. To quantify the errors on our derived parameters we note
that the principal sources of error in the above analysis are (i) shot
noise -- we are inherently limited by the finite number of sources
sampled within a few tidal radii of each lensing cluster galaxy (in
the present analysis we typically have 40 cluster members and $\sim
400$ background sources); (ii) the spread in the intrinsic ellipticity
distribution of the source population; (iii) the unknown source
redshifts and (iv) observational errors arising from uncertainties in
the measurement of ellipticities from the images for the faintest
objects. From simulations where the detailed mass distribution of
clusters is known, we have calibrated the sources of error and find
that of the above-mentioned four factors, shot noise (accounts
typically for $\sim 50\%$ of the incurred error) and the unknown 
redshifts for individual background galaxies (contributes
$\sim 0$--30\% depending on the redshift of the cluster) dominate the
error budget. The remaining $\sim 20\%$ of the error budget arises due 
to inaccuracies in the measurement of shapes. 

We show in Fig.~1 the likelihood contours for the galaxy perturber
models of each of the five clusters.  In all cases we detect an
unambiguous galaxy-galaxy lensing signal at the $>$3-$\sigma$ level --
confirming the existence of truncated dark halos associated with
early-type galaxies in clusters. The likelihood analysis yields
best-fit model parameters: $\sigma^\ast$ the central velocity dispersion
and truncation radius $r_t^\ast$ for a typical $L^\ast$ cluster member,
these values are listed in Table~1.\footnote{The mass obtained for a
typical bright cluster galaxy by Tyson et al.\ (1998) using only
strong lensing constraints inside the Einstein radius of the cluster
Cl\,0024+16 is consistent with our results.}  The mass-to-light ratios
quoted here take passive evolution of the stellar content of
elliptical/S0 galaxies into account modeled using the stellar
population synthesis models of Bruzual \& Charlot (1993).

\section{Discussion and Conclusions}

We have statistically extracted characteristic parameters for typical
$L^\ast$ cluster galaxies that inhabit massive, dense lensing
cluster-lenses ranging in redshift from 0.17--0.58. This has been
achieved by combining strong and weak lensing {\it HST} observations
in conjunction with an assumed parametric mass model.  We find that
the inferred mass distribution of a fiducial $L^\ast$ is extremely
compact, although the inferred $r_t^\ast$'s lie well outside the
optical radii and correspond to roughly between 5--$10\,R_e$.  Our
analysis also shows that the halos of individual cluster galaxies
contribute at most 10--20\% of the total mass of the cluster within
the central 1\,Mpc, covered by the {\it HST} {\it WFPC2} imaging using
the results of our likelihood analysis alongwith the best-fit
paramters that characterize the smooth clump. Therefore, in the inner
regions of these clusters the bulk of the dark matter is in fact
smoothly distributed. Similar lensing studies of field galaxies, e.g.\
Wilson et al.\ (2001), typically find a non-zero signal for the
radially averaged stacked tangential shear out to 200\,kpc. In
contrast our study of the halos of galaxies in clusters detects a
finite $r_t^\ast$, which we attribute to the tidal truncation induced
by the motion of these cluster galaxies inside the potential
well. From the contours in the likelihood plots, the presence of field
galaxy scale dark halos can, in fact, be excluded at $>10$-$\sigma$
significance.

The clusters we study here are all rich systems spanning a range in
central density, which may explain why the best-fit values of
$r_t^\ast$ obtained vary by a factor of 2--3.  To test this suggestion
we plot in Fig.~2 the variation of the central density of the cluster 
dark matter with $r_t^\ast/\sigma^\ast$ based on our lens models
and evaluated at the cluster core radius.  We see a good correlation
and derive a best-fit slope of $-0.6 \pm 0.2$. This compares well with
the theoretical expected value from a tidal stripping model (Merritt 
1983) of $-0.5$:
\begin{eqnarray}
r_t^\ast\,\,\approx\,\,40\,\,(\frac {\sigma_{*}}{180{\rm \,km\,s}^{-1}})\,
({\frac {\rho_0(r_c)}{3.95 \times {10^6}\,M_\odot{\rm\,kpc}^{-3}}})^{-\frac{1}{2}}\,\,{\rm kpc}.
\end{eqnarray}

Dark halos of the scale detected here indicate a high
probability of galaxy encounters over a Hubble time within a rich
cluster.  However, since the internal velocity dispersions of these
cluster galaxies ($<250$\,km\,s$^{-1}$) are much smaller than
their orbital velocities, these interactions are unlikely to lead to
mergers, suggesting that the encounters of the kind simulated in the
`galaxy harassment' picture (Moore et al.\ 1996) are frequent and
likely.  In fact, high resolution cosmological N-body simulations of
cluster formation and evolution (Ghigna et al.\ 1998; Moore et al.\
1996), find that the dominant interactions are between the global
cluster tidal field and individual galaxies after $z = 2$. The cluster
tidal field significantly tidally strips galaxy halos in the inner
$0.5$\,Mpc and the radial extent of the surviving halos is a strong
function of their distance from the cluster center. Much of this
modification is found to occur between $z = 0.5$--0. Detailed
comparison of these results with tidal stripping of dark matter
halos in cosmological N-body simulations will be presented in
a forthcoming paper.   
 
The prospects for extending this technique to larger scales within
clusters in order to study the efficiency of halo stripping as a
function of radius (variation of $r_t^\ast$ as a function of radius) and
morphological type are very promising with new instruments such as the
{\it Advanced Camera for Survey} on {\it HST}. Multi-band imaging will
enable photometric redshift determination for the background sources
which will reduce one of the significant sources of noise for future
analyses.

\acknowledgments

PN acknowledges support from Trinity College, Cambridge for a
Research Fellowship, JPK from the CNRS and the TMR-Lensing collaboration 
and IRS from the Royal Society and the Leverhulme Trust.

\begin{figure*}[tbh]
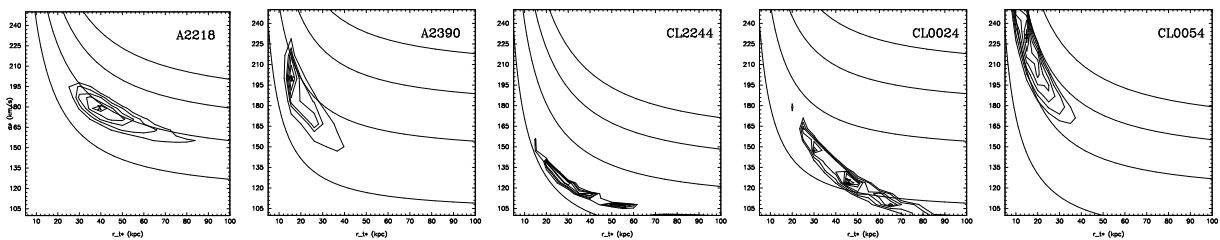

\centerline{\psfig{file=fig1a_nks.ps,width=1.2in,angle=0}
\hspace*{1mm}\psfig{file=fig1b_nks.ps,width=1.2in,angle=0}  
\hspace*{1mm}\psfig{file=fig1c_nks.ps,width=1.2in,angle=0}
\hspace*{1mm}\psfig{file=fig1d_nks.ps,width=1.2in,angle=0}
\hspace*{1mm}\psfig{file=fig1e_nks.ps,width=1.2in,angle=0}}
 \caption{The results of the maximum-likelihood analysis for the
{\it HST} cluster-lenses in our sample, $\sigma^\ast$ and
$r_t^\ast$ correspond respectively to the central velocity dispersion and
outer scale radius (identified as the tidally truncated radius) for a
fiducial $L^\ast$ cluster galaxy in each of these clusters. The contours 
start at 1-$\sigma$ and increase in 1-$\sigma$ increments from inside 
out in all 5 panels. The thick solid lines represent the
contours of constant aperture mass and the best-fit values at
marked by the cross.}
 \end{figure*}

\begin{figure}
\centerline{\psfig{file=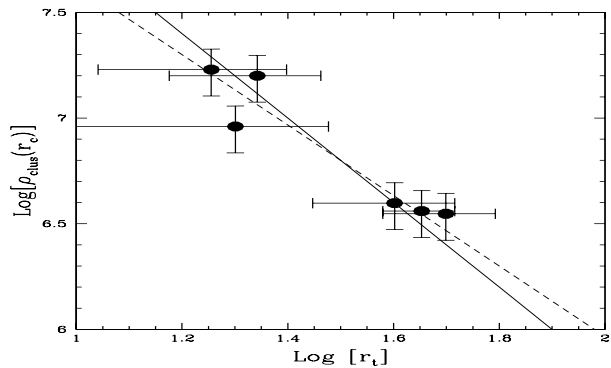,width=3.3in,height=1.98in}}
\caption{Scaling out the variation in the $\sigma^*$'s (as per eqn. 
4 above), the central density of the cluster evaluated at the core 
radius is plotted against the tidal radius. The errors plotted
in $r_t$ are the 3-$\sigma$ values. Performing a least-squares fit to the 
data points, the value of the best-fit power-law index $\eta$, 
where $r_t^* \propto \rho^{\eta}$, is estimated to be $-0.6 \pm 0.2$, 
(dashed line) in excellent agreement with theoretical expectations of 
$\eta = -0.5$ (solid line).}
\end{figure}

\begin{table*}
\begin{center}
\begin{tabular}{lcccccccccc}
\hline\hline\noalign{\smallskip}
${\rm Cluster}$&${z}$&${L_X}$&{T$_{\rm exp}$}&
${\sigma^\ast}$&${r_t^\ast}$&${M_{\rm ap}/L_v}$&$\rm {M^\ast}$&
$\sigma_{clus}$ & ${r_c}$ & ${\rho_{clus}(r_c)}$\\
& & ($10^{44}$\,ergs\,s$^{-1}$) & (ks) & (km\,s$^{-1}$) & (kpc) & 
(M$_\odot$/L$_\odot$) & (10$^{11}$M$_\odot$) & (km\,s$^{-1}$) & (kpc) 
& (10$^6$ $\msun$ kpc$^{-3}$) \\
\noalign{\smallskip}
\hline
\noalign{\smallskip}
{A\,2218} & ${0.17}$ &  9.5 & ~6.3 & ${180\pm10}$ & ${40\pm12}$ &
${5.8\pm1.5}$ & $\sim\,14 $ & ${1070\pm70}$  & ${75\pm10}$ & {3.95}\\

{A\,2390} & ${0.23}$ & 23.4 & 10.5& ${200\pm15}$ & ${18\pm5}$ &
${4.2\pm1.3}$ & $\sim\,6.4 $  &${1100\pm80}$& ${55\pm10}$ & {16.95}\\

{AC\,114}  & ${0.31}$& 13.4 & 16.8 & ${192\pm35}$ & ${17\pm5}$ &
${6.2\pm1.4}$ & $\sim\,4.9 $ &${950\pm50}$& ${45\pm15}$ & {9.12}\\ 

{Cl\,2244$-$02} & ${0.33}$& 4.8 & 10.5 & ${110\pm7}$ & ${55\pm12}$ &
${3.2\pm1.2}$ & $\sim\,6.8 $ &${600\pm80}$ &${30\pm15}$ & {3.52}\\

{Cl\,0024+16} & ${0.39}$&  2.4 & 13.2 & ${125\pm7}$ & ${45\pm5}$ &
${2.5\pm1.2}$ & $\sim\,6.3 $ &${1000\pm70}$ &${30\pm10}$ & {3.63}\\

{Cl\,0054$-$27} & ${0.58}$& 0.25 & 16.8 & ${230\pm18}$ & ${20\pm7}$ &
${5.2\pm1.4}$ & $\sim\,9.4 $ &${1100\pm100}$ &${30\pm10}$ & {15.84}\\
\noalign{\smallskip}
\hline
\end{tabular}
\end{center}
\end{table*}

\end{document}